\documentclass[aps,preprint,showpacs]{revtex4}
\usepackage{amssymb}
\usepackage{amsfonts}
\usepackage{amsmath}
\usepackage{graphicx}

\begin{document}

\title{Matter-wave solitons with the minimum number of particles in
two-dimensional quasiperiodic potentials}
\author{Gennadiy Burlak$^{1}$ and Boris A. Malomed$^{2}$}
\affiliation{$^{1}$Centro de Investigaci\'{o}n en Ingenier\'{\i}a y Ciencias Aplicadas,
Universidad Aut\'{o}noma del Estado de Morelos, Cuernavaca, Mor., M\'{e}xico}
\affiliation{$^{2}$Department of Physical Electronics, School of Electric Engineering,
Faculty of Engineering, Tel Aviv University, Tel Aviv 69978, Israel}

\begin{abstract}
We report results of systematic numerical studies of 2D matter-wave soliton
families supported by an external potential, in a vicinity of the junction
between stable and unstable branches of the families, where the norm of the
solution attains a minimum, facilitating the creation of the soliton. The
model is based on the Gross-Pitaevskii equation for the self-attractive
condensate loaded into a quasiperiodic (QP) optical lattice (OL). The same
model applies to spatial optical solitons in QP photonic crystals. Dynamical
properties and stability of the solitons are analyzed with respect to
variations of the depth and wavenumber of the OL. In particular, it is found
that the single-peak solitons are stable or not in exact accordance with the
Vakhitov-Kolokolov (VK) criterion, while double-peak solitons, which are
found if the OL wavenumber is small enough, are always unstable against
splitting.
\end{abstract}

\maketitle

\textit{Introduction and the model}. A challenging subject in studies of
dynamical patterns in Bose-Einstein condensates (BECs) and nonlinear optics
is the creation of matter-wave or photonic solitons in multidimensional
settings \cite{Review,Moti,Review2}. Various routes to the making of stable
two- and three-dimensional (2D and 3D) fundamental and vortical solitons
have been elaborated theoretically. As demonstrated in Refs. \cite{BBB}-\cite%
{BBB3}, universal stabilization methods for the matter-wave and optical
solitons are provided, respectively, by optical lattices (OLs) or photonic
crystals, i.e., essentially, by spatially periodic potentials. OLs are
induced, as interference patterns, by coherent laser beams illuminating the
condensate in opposite directions, while photonic lattices may be created,
by means of various technologies, as permanent structures in optical
waveguides, or as virtual photoinduced structures in photorefractive
crystals \cite{Moti}. A more difficult but also realistic possibility is
stabilizing solitons by means of nonlinear lattices, i.e., spatially
periodic modulations of the nonlinearity coefficient \cite{Review2}. In
principle, similar methods may be applied to a gas of polaritons \cite%
{polariton}, where the evidence of the BEC state was reported too \cite%
{Kasprzak:2006a}, using properly engineered superlattices \cite{devices}.

The stabilization of 2D and 3D solitons is possible with the help of the
fully-dimensional OL, whose dimension is equal to that of the entire space, $%
D$, and by low-dimensional lattices, with dimension $D-1$ \cite%
{BBB2,Barcelona}, \cite{Fatkhulla}. Other methods for the creation of robust
solitons rely on the time-periodic \textit{management} \cite{Malomed:2006a}
of nonlinear \cite{FRM,FRM-in-trap,Warsaw} or linear \cite{trap-modulation}
characteristics of the condensate (following the method proposed \cite{Isaac}
and later implemented experimentally \cite{Centurion} for the stabilization
of 2D solitons in optics by means of the periodic modulation of the Kerr
coefficient along the propagation distance ). In these contexts, the
stability of the matter waves in 2D OLs, and under various scenarios of the
time-periodic management, has been studied extensively, see, e.g., Refs.
\cite{Montesinos:2004a,Burlak}. In addition, the stabilization of multidimensional solitons may be
provided by nonlocal (dipole-dipole) interaction between atoms \cite%
{Pedri:2005a} or nonlocal (thermal) nonlinearity in optics \cite{thermal}.

Besides periodic OLs, quasiperiodic (QP) ones have also drawn a great deal
of interest---in particular, as the simplest setting for the realization of
the Anderson localization of matter waves \cite{Anderson}. The self-trapping
of 2D solitons in QP potentials was studied too \cite%
{BBB3,HidetsuguSakaguchi:2006a,Gena}. The objective of this work is to
extend the previously reported analysis of the stabilization of 2D solitons
by lattice potentials to the case of QP lattices and self-attractive
nonlinearity (negative scattering length of inter-atomic interactions in the
BEC), which can be readily implemented in $^{7}$Li and $^{85}$Rb condensates
\cite{experiment}, and corresponds to the usual Kerr nonlinearity in optics.
As known from the previous analyses \cite{BBB,BBB2,HidetsuguSakaguchi:2006a}%
, the dependence between the chemical potential and the norm (which is
proportional to the number of atoms in BEC, or total power of the optical
beam) for 2D solitons supported by lattice potentials, $\mu (N)$, features
two branches, stable and unstable ones [with $d\mu /dN<0$ and $d\mu /dN>0$,
respectively, according to the Vakhitov-Kolokolov (VK) criterion \cite{VK}].
The branches merge at a threshold (minimal) value of $N$, below which the
solitons decay due to the delocalization transition \cite{Salerno}.

Our analysis is based on the 2D Gross-Pitaevskii equation for the BEC\
mean-field wave function, $\Psi\left( x,y,t\right) $, written in the
dimensionless form assuming the self-attractive nonlinearity \cite{BEC}:
\begin{equation}
i\frac{\partial\Psi}{\partial t}+\frac{1}{2}\left( \frac{\partial^{2}}{%
\partial x^{2}}+\frac{\partial^{2}}{\partial y^{2}}\right) \Psi+\left\vert
\Psi\right\vert ^{2}\Psi+V(x,y)\Psi=0,  \label{GPE}
\end{equation}
where the QP lattice potential of depth $2V_{0}$ is taken as \cite%
{BBB3,HidetsuguSakaguchi:2006a,Gena}

\begin{equation}
-V(x,y)=-V_{0}\sum_{n=1}^{M}\cos (\mathbf{k}^{(n)}\mathbf{r)}\text{,}
\label{OptLatticeV}
\end{equation}%
with the set of wave vectors $\mathbf{k}^{(n)}=k\{\cos \left( 2\pi
(n-1)/M\right) ,\sin \left( 2\pi (n-1)/M\right) \}$ and $M=5$ or $M\geq 7$.
Here, following Ref. \cite{HidetsuguSakaguchi:2006a}, we focus on the basic
case of the Penrose-tiling potential, corresponding to $M=5$. The 2D profile
of the potential is displayed below in Fig. \ref{Pic_Fig3}(d). Setting $%
V_{0}>0$, the center of the 2D soliton will be placed at the local minimum
of potential (\ref{OptLatticeV}), $x=y=0$. The solitons will be
characterized by their norm, defined as usual: $N=\int \int \left\vert \Psi
(x,y)\right\vert ^{2}dxdy$. The relation of $N$ to the actual number of
atoms in the condensate, $\mathcal{N}$, is given by means of standard rescaling \cite%
{BEC}: $\mathcal{N}=\left( a_{\perp }/4\pi a_{s}\right) N$, where $a_{\perp
} $ (typically, $\sim \mathrm{\mu }$m) and $a_{s}$ ($\sim 0.1$ nm) are the
transverse trapping length of the condensate and scattering length of the
atomic collisions, respectively. In optics, the same equation (\ref{GPE}),
with $t$ replaced by the propagation distance, $z$, governs, the transmission
of electromagnetic waves with local amplitude $\Psi $ in the bulk waveguide
with the transverse QP modulation of the refractive index. In the latter
case, $N$ is proportional to the beam's total power.

\textit{Numerical results: soliton families}. Simulations of Eq. (\ref{GPE})
were performed on the 2D numerical grid of size $128\times128$, starting
with the input in the form of an isotropic Gaussian,
\begin{equation}
\Psi(x,y)=A_{0}\exp(-q(x^{2}+y^{2})).  \label{initial}
\end{equation}
Initial amplitude $A_{0}$, along with the OL depth and wavenumber, $V_{0}$
and $k$, were varied, while the initial width was fixed by setting $q=0.9$
[which is possible by means of rescaling of Eq. (\ref{GPE})].

Before proceeding to numerical results, it is relevant to note that,
although the application of the variational approximation, which is a
ubiquitous analytical tool for the study of bound states in nonlinear
systems \cite{Review,Review2}, to 2D solitons in QP potentials is possible
\cite{BBB3}, the simplest isotropic \textit{ansatz}, taken in the same form
as Gaussian (\ref{initial}), cannot capture peculiarities of the setting
based on the QP potential. Indeed, the part of the Lagrangian accounting for
the interaction of ansatz (\ref{initial}) with the underlying OL potential (%
\ref{OptLatticeV}) consists of integrals like $V_{0}A_{0}^{2}\int \int \cos
\left( \mathbf{k}^{(n)}\mathbf{r}\right) \exp (-2qr^{2})\mathbf{dr}=\pi %
\left[ V_{0}/\left( 2q\right) \right] \exp \left[ -k^{2}/\left( 8q\right) %
\right] $. Being insensitive to the particular orientation of wave vectors $%
\mathbf{k}^{(n)}$, this approximation is too coarse. It may be improved by
using an anisotropic ansatz, but this will render the variational analysis
cumbersome.

The first objective is to construct families of localized ground-state
modes, in the form of $\Psi (x,y,t)=\exp (-i\mu t)\varphi (x,y)$, with real
wave function $\varphi (x,y)$\ found by means of the accelerated
imaginary-time method \cite{Yang:2008a}. Following the convention commonly
adopted in physics literature \cite{Review}-\cite{BBB3}, \cite{Barcelona}-%
\cite{Warsaw}, we refer to these modes as \textquotedblleft solitons", even
though they do not feature the unhindered motion characteristic to
\textquotedblleft genuine" solitons. The simulations of Eq. (\ref{GPE}),
rewritten in the imaginary time with a fixed value of $\mu $, quickly
converge to the ground state, with $\lesssim 1000$ iterations necessary to
reduce the residual error to the level of $10^{-10}$.

\begin{figure}[tbp]
\begin{center}
\includegraphics[
height=4.2434in,
width=5.844in
]{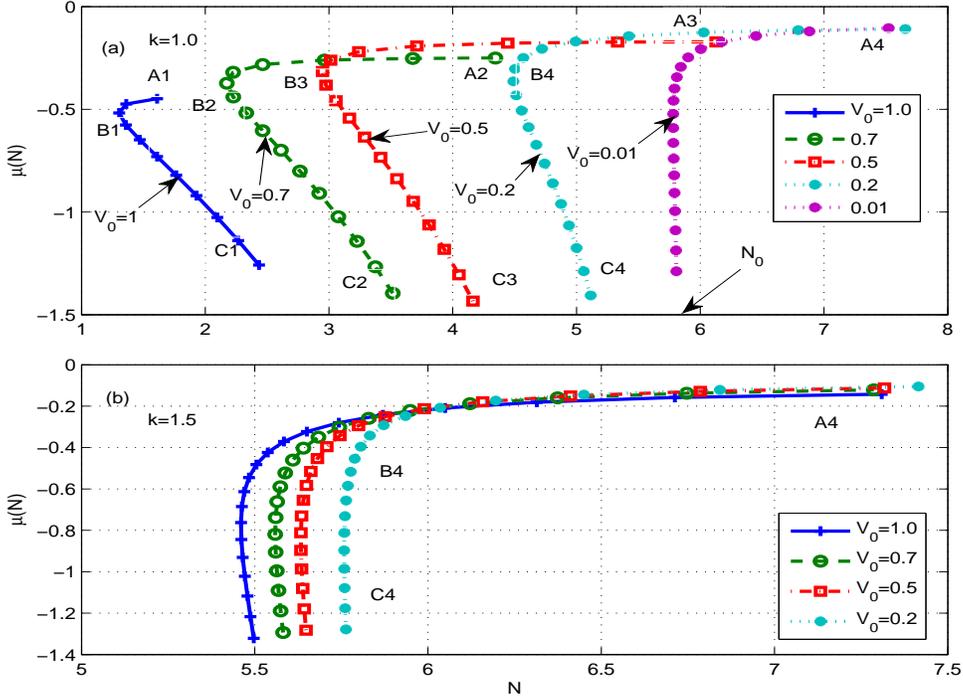}
\end{center}
\caption{(Color online) Chemical potential $\protect\mu $ of the
ground-state mode (\textquotedblleft soliton") versus its norm $N$, at two
fixed wavenumbers of the Penrose-tiling potential, $k=1$ (a) and $k=1.5$
(b), and different values of its depth, $V_{0}$. Labels $Cj$ and $Aj$ (%
$j=1,2,3,4$) indicate VK-stable and unstable branches, respectively, while
points $Bj$ mark junctions between them.}
\label{Pic_Fig1}
\end{figure}

\begin{figure}[tbp]
\begin{center}
\includegraphics[
height=3.9851in,
width=5.8885in
]{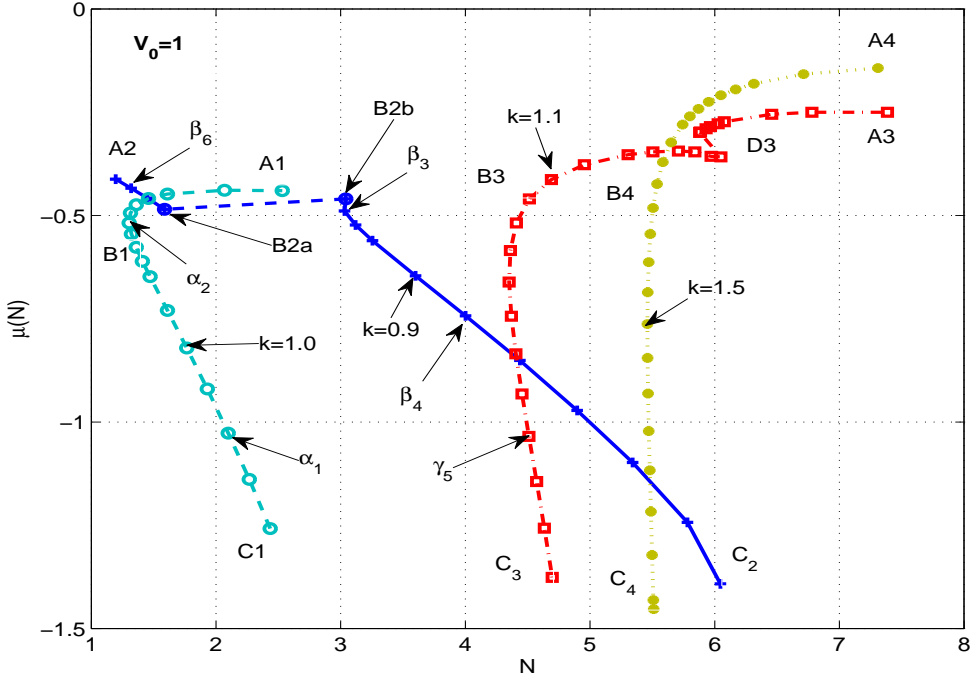}
\end{center}
\caption{(Color online) The same as in Fig. \protect\ref{Pic_Fig1}, but for
the fixed depth of the OL potential, and different values of its wavenumber.}
\label{Pic_Fig2}
\end{figure}

In Fig. \ref{Pic_Fig1}, chemical potential $\mu $ of the ground state is
shown, as a function of its norm $N$, at two fixed wavenumbers of the
Penrose-tiling potential, $k=1$ (a) and $k=1.5$ (b) and various values of
its depth, $V_{0}$. Further, Fig. \ref{Pic_Fig2} shows $\mu (N)$ for fixed $%
V_{0}$ and different values of $k$. Labels $\mathrm{Cj}$ and $\mathrm{Aj}$ ($%
\mathrm{j}=1,2,3,4$) indicate branches which are expected to be stable and
unstable according to the Vakhitov-Kolokolov (VK) criterion \cite{VK,Berge'}%
, i.e., with $d\mu /dN<0$ and $d\mu /dN>0$, respectively. The imaginary-time
algorithm, which generated the solitons, ceased to converge at lower
termination points of the branches shown in Figs. \ref{Pic_Fig1} and \ref%
{Pic_Fig2}, where the amplitude of the solution becomes too large.

Points $\mathrm{Bj}$ in Figs. \ref{Pic_Fig1} and \ref{Pic_Fig2} mark the
\textit{junctions} between the stable and unstable branches, where $d\mu /dN$
diverges, while $N$ attains its minimum. At small $V_{0}$ [see the curve for
$V_{0}=0.01$ in Fig. \ref{Pic_Fig1}(a)], the values of $N$ on the VK-stable
branches approach the limit value, $N_{\mathrm{Townes}}\approx 5.85$, which
corresponds to the \textit{Townes soliton} in the free 2D space \cite{Berge'}%
.

As said above, the main point in this work is the study of the solitons
close to norm-minimizing points $\mathrm{Bj}$, which are of obvious interest
to the potential experiment. In Fig. \ref{Pic_Fig1} we observe that stable
solitons with the minimum norm (i.e., smallest number of atoms) are,
naturally, generated in the deepest potential, represented by families $%
\mathrm{A1-B1-C1}$. The norm attains its minimum, $N_{\min }=1.304$ (with $%
\mu =-0.518$) at $k=1$ and $V_{0}=1$ [point $\mathrm{B1}$ in Fig. \ref%
{Pic_Fig1}(a)]. We also observe that the stability range (the distance from
the lower termination point to point $\mathrm{B1}$) in Fig. \ref{Pic_Fig1}%
(a) for $k=1$ is $\Delta N=N(\mathrm{C1})-N(\mathrm{B1})=2.531-1.304=%
\allowbreak 1.\,\allowbreak 227$, which is $\simeq 20$ times larger than $%
\Delta N=5.510-5.460=\allowbreak 0.05$ $\,$in Fig. \ref{Pic_Fig1}(b) for $%
k=1.5$, at\ the same OL depth, $V_{0}=1$. Generally, the comparison of Fig. %
\ref{Pic_Fig1}(a) and Fig. \ref{Pic_Fig1}(b) demonstrates that, for given $%
V_{0}$, the norm of the ground states strongly depends on the OL wavenumber,
$k$.

The branch $\mu(N)$ with $k=0.9$ in Fig. \ref{Pic_Fig2} is notably different
from other branches with $k\geq1$. Although a continuous dependence $\mu(N)$
is found in the range of $\mathrm{C2-B2b}$, no solutions have been found
(the imaginary-time algorithm does not converge to them) between points%
\textrm{\ }$\mathrm{B2b}$ and $\mathrm{B2a}$ (the dashed segment $\mathrm{%
B2b-B2a}$ is depicted in Fig. \ref{Pic_Fig2} only as a guide to the eye).
The algorithm again converges to the ground-state modes in the range of $%
\mathrm{B2a-A2}$.

\begin{figure}[tbp]
\begin{center}
\includegraphics[
height=4.38in,
width=6.0328in
]{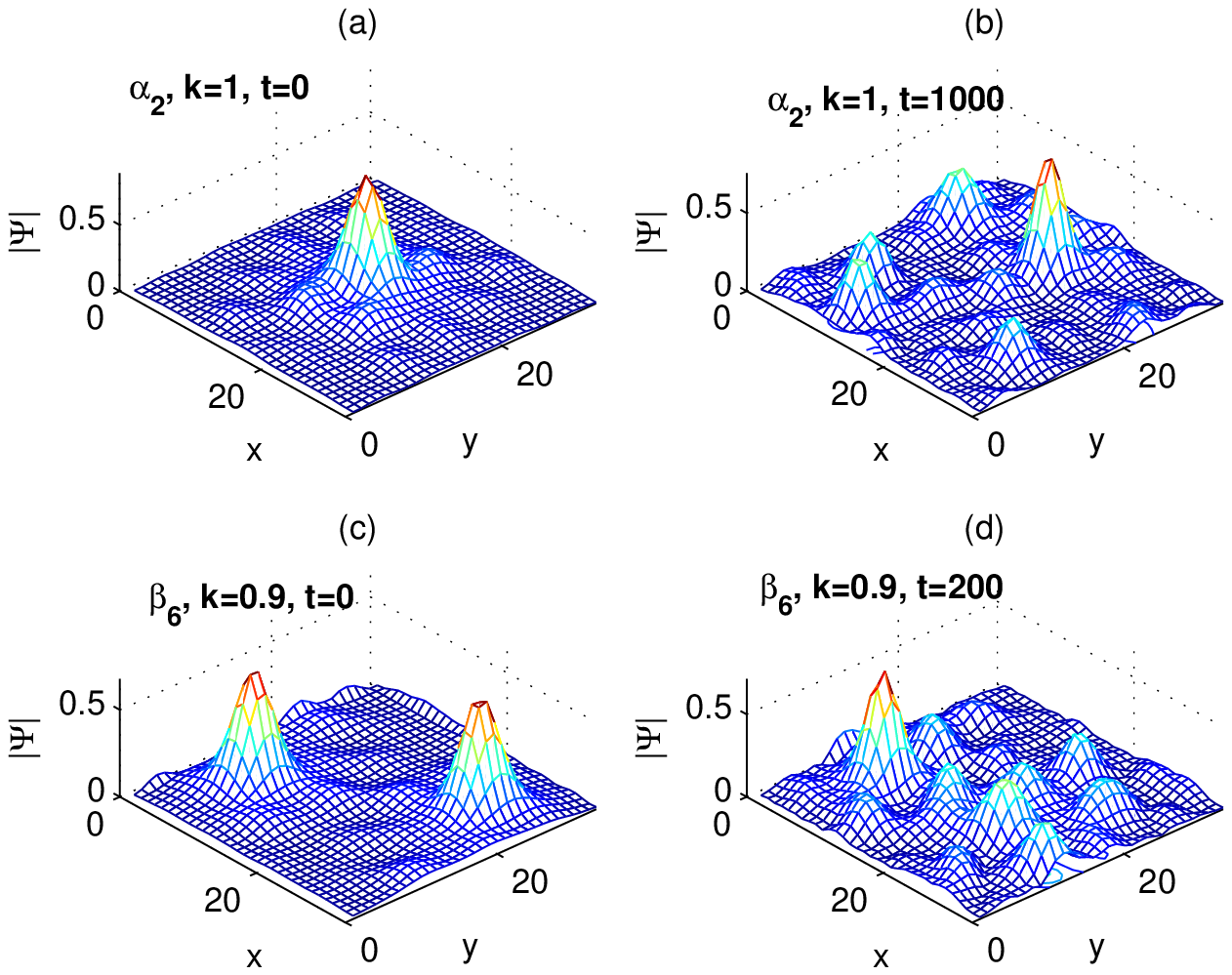}
\end{center}
\caption{(Color online) The evolution of the stationary modes labeled by
points $\protect\alpha _{2}$ and $\protect\beta _{6}$ in Fig. \protect\ref%
{Pic_Fig2}, for $k=0.9$ and $V_{0}=1$. (a) The shape of mode $\protect\alpha %
_{2}$, with $N=1.304$, $\protect\mu =-0.518$; (b) the result of the
evolution at $t=1000$. The final state is not a bound one, but it does not
decay into radiation. (c,d) Mode $\protect\beta _{6}$\ with $N=1.319$, $%
\protect\mu =-0.434$, which splits into two parts, and eventually decays.}
\label{Pic_Fig4}
\end{figure}

\begin{figure}[ptb]
\centering
\includegraphics[
height=4.4227in, width=5.8807in
]{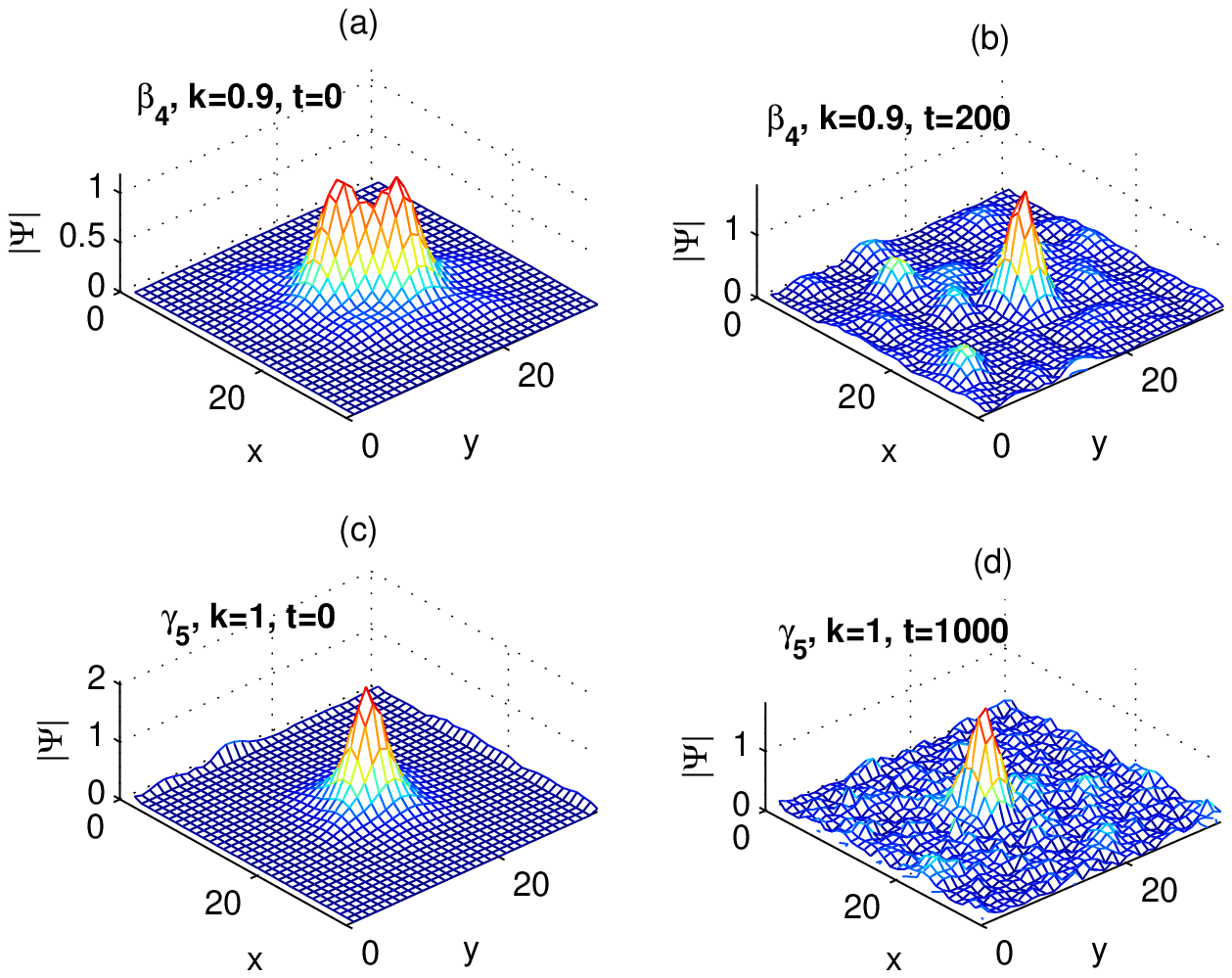}
\caption{(Color online) The perturbed evolution of modes $\protect\beta_{4}$
(a,b) and $\protect\gamma_{5}$ (c,d), which are marked in Fig. \protect\ref%
{Pic_Fig2}). The former one, with $N=4.002$ and $\protect\mu=-0.743$, is
unstable, while the latter mode ($\protect\gamma_{5}$), with $N=4.510$ and $%
\protect\mu=-1.035$, is stable.}
\label{Pic_Fig5}
\end{figure}

Furthermore, a tail of segment $\mathrm{B2b-C2}$ of this branch penetrates
into the overcritical region, $N=6.046>N_{\mathrm{Townes}}=5.85$. This
feature is explained by the fact that the solitons found at $k\leq 0.9$ (in
particular, the ones marked by $\beta _{3},\beta _{4},\beta _{6}$ in Fig. %
\ref{Pic_Fig2}) are actually double-humped structures, featuring pairs of
spatially separated or almost fused density peaks [see Figs. \ref{Pic_Fig4}%
(c) and \ref{Pic_Fig5}(a), respectively].

\textit{Stability of the solitons}. The VK criterion does not guarantee the
full stability of solitons, as it does not capture instabilities associated
with complex eigenvalues. To test the full stability, we simulated perturbed
evolution of the solitons over a sufficiently long interval, typically $%
t=1000$ (which covers, roughly, $10$ diffraction times of the corresponding
localized states), adding small random perturbation to the initial
conditions, with a relative amplitude $\sim 0.01$. The modes whose evolution
was tested in this way are indicated by arrows in Fig. \ref{Pic_Fig2},
attached to symbols $\alpha _{1},\alpha _{2}$ and $\beta _{3},\beta
_{4},\beta _{6}$, which pertain to branches with $k=1$ and $k=0.9$,
respectively, and $\gamma _{5}$, that pertains to $k=1.1$. The results of
the evolution simulations are shown in Figs. \ref{Pic_Fig3}-\ref{Pic_Fig5}.

Figure \ref{Pic_Fig3} presents details of the stability test for the ground
state on branch $\mathrm{C1}$, marked by $\alpha _{1}$ in Fig. \ref{Pic_Fig2}
(for $V_{0}=1$ and $k=1$), with the norm and chemical potential $N=2.098$
and $\mu =-1.027$. This mode is stable.\textbf{\ }

\begin{figure}[tbp]
\begin{center}
\includegraphics[
height=5.1189in,
width=6.0926in
]{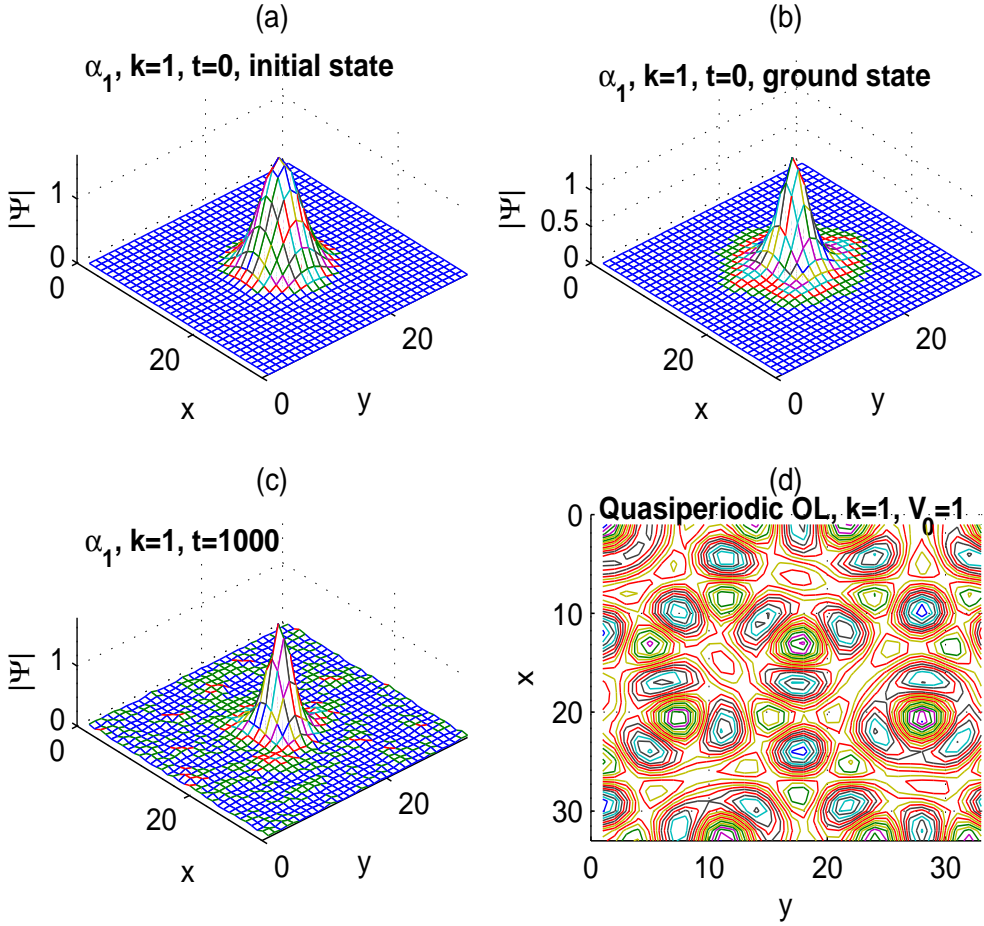}
\end{center}
\caption{(Color online.) The spatial structure of the stable localized mode
supported by the quasiperiodic potential, labeled by $\protect\alpha _{1}$
in Fig. \protect\ref{Pic_Fig2}. (a) The Gaussian initial configuration (%
\protect\ref{initial}) for $V_{0}=1$ and $k=1$, transformed by the
imaginary-time relaxation into the ground state, which is shown in panel
(b). Panel (c): The result of the perturbed evolution (in real time) at $%
t=1000$. (d) The contour-plot profile of the underlying quasiperiodic
potential with $V_{0}=1$ and $k=1.0$.}
\label{Pic_Fig3}
\end{figure}

Figures \ref{Pic_Fig4} (a) and (b) display the evolution of the solitons
taken near the junction points between the VK-stable and unstable segments
of the $\mu (N)$ curves, for $k=1$ and $k=0.9$. Figure \ref{Pic_Fig4}(a)
pertains to the mode labeled $\alpha _{2}$ (with $k=1$) in Fig. \ref%
{Pic_Fig2}, which evolves into the perturbed state depicted at $t=200$ in
Fig. \ref{Pic_Fig4}(b). This mode is unstable, splitting into a set of
density peaks located at different potential minima, which, however, do not
tend to decay into dispersive waves. The evolution of another unstable mode,
labeled by $\beta _{6}$ in Fig. \ref{Pic_Fig2}, is displayed in Figs. \ref%
{Pic_Fig4}(c) and \ref{Pic_Fig4}(d). It splits into two parts, and
eventually decays into the dispersive radiation either.

Finally, Fig. \ref{Pic_Fig5} represents the perturbed evolution of the mode
with larger norms ($N>4$), for $k=0.9$ and $k=1.1$, which correspond to
points $\beta_{4}$ and $\gamma_{5}$, respectively, labeled in Fig. \ref%
{Pic_Fig2}. In panels \ref{Pic_Fig5}(a,b) we again observe that the former
(double-peak) mode, corresponding to $k=0.9$, does not produce a stable
soliton in the course of the perturbed evolution. However, Fig. \ref%
{Pic_Fig5}(d) demonstrates that the soliton corresponding to point $%
\gamma_{5}$ is stable. The eventual conclusion following from the analysis
of the numerical results is that all the double-peak structures are unstable
against splitting, irrespective of their formal compliance with the VK
criterion, while the single-peak solitons are stable or not in the exact
accordance with VK.

\textit{Conclusion}. We have studied the dynamics of 2D matter-wave solitons
near the junction points between the stable and unstable branches of curves $%
\mu (N)$ for the soliton families supported by the interplay of the
self-attractive nonlinearity and Penrose-tiling OL potential. These points
are interesting to physical applications, as they correspond to the smallest
number of atoms which is necessary to build 2D matter-wave solitons, or the
smallest total power necessary for the making of spatial optical solitons.
It was found that the shape and stability of such solitons crucially depend
on the depth and period of the OL. A challenging problem is to extend the
analysis to vortex solitons supported by quasi-periodic potentials\cite{BBB3}.

This work was supported, in a part, by CONACyT/SEP 2012 and PROMEP CA Redes projects (M\'{e}xico).

\end{document}